\documentclass[onecolumn,12pt,a4]{llncs}
\usepackage[T1]{fontenc}
\usepackage{amssymb}
\usepackage{amsmath}
\usepackage{algorithm}
\usepackage{algorithmic}

\newtheorem{lema}{Lemma}[section]

\newenvironment{demo}{\noindent{\bf Proof:}}{\\\hspace*{\fill}\bloco}

\newcommand{\bloco}{\hfill\rule{2mm}{2mm}}
\newcommand{\al}{\alpha}

\def\sVEV#1{\left\langle #1\right\rangle}     % variable < >
\def\sket#1{\left| #1\right\rangle}           % variable | >

\def\Z{\mathbb{Z}}

\def\up{\textup}
\floatname{algorithm}{Algorithm}

\title{Solution to the Hidden Subgroup Problem for a Class of Noncommutative Groups}

\author{Demerson N. Gon\c{c}alves\inst{1,2}, Renato Portugal \inst{2}, }
\institute{Universidade Cat\'{o}lica de Petr\'{o}polis - CEC/UCP,
   25685-070, Petr\'{o}polis, RJ \and
  Laborat\'{o}rio Nacional de Computa\c{c}\~{a}o Cient\'{\i}fica - LNCC,
         25651-075,  Petr\'{o}polis, RJ \\
  demerson.goncalves@ucp.br, portugal@lncc.br}

\begin{document}

\maketitle

\begin{abstract}
The hidden subgroup problem (HSP) plays an important role in quantum computation, because many quantum algorithms that are exponentially faster than classical algorithms can be casted in the HSP structure. In this paper, we present a new polynomial-time quantum algorithm that solves the HSP over the group $\Z_{p^r} \rtimes \Z_{q^s}$, when $p^r/q= \up{poly}(\log p^r)$, where $p$, $q$ are any odd
prime numbers and $r, s$ are any positive integers.  To find the hidden subgroup, our algorithm uses the abelian quantum Fourier transform and a reduction procedure that simplifies the problem to find cyclic subgroups.
\end{abstract}

\section{Introduction}

The area of quantum algorithms is reviewed in two noteworthy papers~\cite{childs,michele}. A wide class of algorithms deals with algebraic problems~\cite{childs} and most of them can be casted as a Hidden Subgroup Problem (HSP)~\cite{cris}. The HSP can be described as follows: given a group $G$ and a coset-injective function $f:G \rightarrow X$ on some set $X$ such that $f(x)=f(y)$ iff $x\cdot H=y\cdot H$ for some subgroup $H$, the problem consists in determining a generating set for $H$ by querying function $f$. We say that the function $f$ hides the subgroup $H$ in $G$ or $f$ separates the cosets of $H$ in $G$. A quantum algorithm for the HSP is said to be efficient when its computational complexity is polylogarithm in the order of the group, \textit{i.e.} $O(\up{poly}(\log |G|))$. There are many examples of efficient quantum algorithms for the HSP~\cite{simon94,shor2}.  It is known that for finite abelian groups, the HSP can be solved efficiently in a quantum computer~\cite{cris}.  On the other hand, it is not known an efficient solution for a generic nonabelian group.  Two important groups in this context are the symmetrical and the dihedral groups. An efficient algorithm for solving the first one implies in an efficient solution for the graph isomorphism problem~\cite{Beals} and for the second one solves instances of the problem of finding the smallest vector in a lattice, with applications in cryptography~\cite{Regev1}.

An important strategy to solve the nonabelian HSP combines three methods: (1) the abelian Fourier transform, (2) the characterization of all subgroups of the group $G$, and (3) the quotient group reduction.  This strategy was first employed by Ettinger and H{\o}yer~\cite{Ettinger3}, that have reduced the HSP in the dihedral group to the problem of finding cyclic subgroups of order 2.  Later on, also employing the same strategy, Inui and Le Gall~\cite{legal} presented an efficient quantum algorithm for the HSP in groups of the form $\mathbb{Z}_{p^r}^m\rtimes \mathbb{Z}_{p}$ with prime $p$ and positive integers $r$ and $m$. Bacon et. al.~\cite{bacon1} solved in polynomial time the HSP in groups of the form $\Z_N \rtimes \Z_q$, for positive integers $N$ and prime $q$, such that $N/q=\up{poly}(\log N) $, reducing the problem to find cyclic subgroups of order $q$.  Recently, Gon\c{c}alves et. al.~\cite{DRC1} presented a class of efficient quantum algorithm for the HSP in $\Z_{p} \rtimes \Z_{q^s}$, with $p/q = \up{poly}( \log p)$, where $p$, $q$ are distinct odd prime numbers and $s$ an arbitrary positive integer. For an extensive description of solutions to the HSP, methods, and references, see Refs.~\cite{wang,cris}.

In this work, we present a new quantum algorithm in polynomial time that solves the HSP in $\Z_{p^r} \rtimes \Z_{q^s}$, with $p^r/q = \up{poly}( \log p^r)$, where $p$, $q$ are distinct odd prime numbers and $r, s$ arbitrary positive integers.  This result extends the results obtained in Ref.~\cite{DRC1}, that are reproduced when $r=1$.   This work generalizes one of the results of Ref.~\cite{MRRS}, which works for prime $N$.   It also generalizes results of Ref.~\cite{MRRS} for $p$-hedral groups and of Ref.~\cite{bacon1}, which are reproduced when $s=1$.

The article is organized as follows.  In Sec.~\ref{possibilidades}, we define the semidirect product $ \Z_{p^r} \rtimes \Z_{q^s}$ and we characterize all subgroups.  In Sec.~\ref{algquant}, we show that the HSP in $ \Z_{p^r} \rtimes \Z_{q^s}$ can be reduced to the problem of finding cyclic subgroups.  In Sec.~\ref{polynomial}, we present a polynomial-time quantum algorithm for the HSP in a class of groups of the form $ \Z_{p^r} \rtimes \Z_{q^s}$.  Finally, in Sec.~\ref{conclusao}, we present our conclusions.

\section{The Structure of the Group $ \Z_{p^r}\rtimes \Z_{q^s}$ }
\label{possibilidades}

Let $p$, $q$ be prime numbers and $r, s$ positive integers. The semidirect product $\Z_{p^r}\rtimes_{\phi}\Z_{q^s}$, where $\Z_{p^r}$ and $\Z_{q^s}$ are cyclic groups and $\phi : \Z_{q^s} \rightarrow \mbox{Aut}(\Z_{p^r} )$, is a group homomorphism that defines the group product. The elements are $(a, b)$, where $a\in \Z_{p^r}$, $b\in \Z_{q^s}$ and the product of two element is $(a,b)(c,d)=(a+\phi(b)(c), b+d)$. Note that $x=(1,0)$ and $y=(0,1)$ generate the group $\Z_{p^r}\rtimes \Z_{q^s}$. Since $\mbox{Aut}(\Z_{p^r})$ is isomorphic to $ \Z_{p^r}^{*}$, the homomorphism $\phi$ is completely determined by $\alpha:=\phi(1)(1)\in \Z_{p^r}^{*}$. The notation $(a,b)$ is equivalent to $x^ay^b$ and the commutation relation is $y^bx^a=x^{a\al^b}y^b$.

Observe that $\phi (0) =\phi (q^s):\Z_ {p^r}\rightarrow\Z_ {p^r} $ is the identity element of the group $\textup {Aut} (\ \Z_ {p^r}) $. Then $\alpha^ {q^s} =\phi (q^s) (1)=1. $ The element $\alpha\in\Z_ {p^r} ^ {*} $ defines the semidirect product of groups $\Z_ {p^r}\rtimes_ {\alpha}\Z_ {q^s} $, if it satisfies the congruence equation
$%\label {congruencia1}
 X^ {q^s}\equiv 1 \,\textup {mod} \, p^r.
$
In this case, we must have $\textup {ord} (\alpha) =q^ {t} $ for some $t=0,\ldots, s$. The case $t=0$ reduces to the direct product of groups $\Z_ {p^r}\times\Z_ {q^s} $, which is an abelian group. An efficient solution for the HSP is known for this case. From now on we consider $1\leq t\leq s$.

For all prime $p$ and positive integer $r$, the group $\Z_ {p^r} ^ {*} $ is cyclic. Let $u\in\Z_ {p^r} ^ {*} $ be an arbitrary generator of this group. Then $\textup {ord} (u)=p^ {r-1} (p-1) $ and $\alpha=u^ {k} $, for some integer $1\leq k < p^ {r-1} (p-1) $. Thus $\alpha^ {q^t} =u^ {kq^t}\equiv \, 1 \,\textup {mod} \, p^r\Rightarrow p-1\mid kq^t. $ Because $p$ and $q$ are distinct prime numbers, we must have $q^t\mid p-1$ and $ k=\frac {lp^ {r-1} (p-1)}{q^t},$ for some $l\in\Z_ {q^t} ^ {*} $. Thus, for each $1\leq t\leq s$ and $l\in\Z_ {q^t} ^ {*} $, the number
\begin {equation}\label{alpha_tl}
\alpha: =u^ {\frac {lp^ {r-1} (p-1)}{q^t}}
\end {equation}
defines a semidirect product of groups, that will be denoted by
\begin{equation}
G_{t,l}=\Z_{p^r}\rtimes_{\al}\Z_{q^s}.
\end{equation}

The parameter $l$ in Eq.~(\ref{alpha_tl}) is disposable, because the group $G_{t,l}$ is isomorphic  to $G_{t,1}$, for all $l$. Next theorem states this fact.
\begin{theorem}
$G_{t,l}\simeq G_{t,1}$ for all $l\in \Z_{q^t}^{*}$.
\end{theorem}
\begin{demo}
Consider the mapping $\Phi_{t,l}:G_{t,1}\rightarrow G_{t,l}$ defined by
$
\Phi_{t,l}(x^ay^b)=x^ay^{l^{-1}b}.
$
Notice that there is an unique inverse $l^{- 1} $ of $l \in \Z_ {q^t} ^ {*} $. Thus, given $x^ay^b \in G_ {t, l} $ there is an unique $x^ay^ {lb} \in G_ {t, 1} $ such that $ \Phi_ {t, l} (x^ay^ {lb}) =x^ay^b$. Then $ \Phi_ {t, l} $ is one-to-one. It is easy to verify that $ \Phi_ {t, l} (x^ay^bx^cy^d) = \Phi_ {t, l} (x^ay^b) \Phi_ {t, l} (x^cy^d) $, therefore $ \Phi_ {t, l} $ is a group isomorphism.
\end{demo}

Let us denote $G_{t,l}$ by $G_t$, where the homomorphism $\al$ is given by Eq.~(\ref{alpha_tl}) with $l=1$. Using the relation $y^bx^a=x^{a\alpha^b}y^b$ and making an induction on $k$, we verify that
\begin{equation}
(x^ay^b)^k=\left\{
\begin{array}{ll}
x^{ak}y^{bk}, & \textup{if} \; q^t\mid b; \\
x^{\frac{a(\alpha^{bk}-1)}{\alpha^b-1}}y^{bk}, & \textup{otherwise},
\end{array}
\right.
\end{equation}
where $q^t=\up{ord}(\alpha)$.

Now we are able to list all subgroups of $G_t= \Z_{p^r}\rtimes \Z_{q^s}$ by stating the following
\begin{theorem}\label{teo-classif}
The subgroups of $G_t$ are
\begin{description}
\item [i)] $\sVEV{x^{p^i}y^{q^j} }$ for $0\leq i \leq r$ and $t\leq j \leq s$;
\item [ii)]$\sVEV{x^{p^i},x^{a}y^{q^j}}$ for $0\leq i \leq r$, $0\leq a < p^r$ and
$0\leq j < t$.
%
%\item [iii)] $\sVEV{x^{i}y^{q^j},x^ay^{q^k}}$ para $0\leq i \leq r$, $t\leq j \leq s$, $0\leq a < p^r$ e $0\leq k < t$.
\end{description}
\end{theorem}
\begin{demo}
In the appendix.
\end{demo}
\section{The Quantum Algorithm}
\label{algquant}

In this section, we show that the HSP in $G_t$ can be reduced to
the problem of finding cyclic subgroups of the form
$\sVEV{x^ay^{q^j}}$, where $a$ is an arbitrary element in the
cyclic group $\Z_{p^r}$ and $0\leq j <t$. Afterward we present an
efficient quantum algorithm for the HSP in $G_t$ for $t=1$.

Let $f$ be the oracle function that hides the subgroup $H$ in $G_t$.  It follows from Theorem~\ref{teo-classif} that there are two cases for $H$, either $\sVEV{x^{p^i}y^{q^j}}$ or $\sVEV{x^{p^i},x^{a}y^{q^j}}$. The parameters to be determined are $i, j$ and $a$.  Parameter $a$ is the most difficult one to address. The algorithm that determines the value of $a$ will be presented in Sec.~\ref{polynomial}.

The general idea of the algorithm is the following.  Let $H_{x}=H\cap \sVEV{x}$ and $H_y=H\cap \sVEV{y}$.  Consider function $f_x$ defined by $ f_x(a)=f(a,0),$ which hides $H_x$ in $ \Z_{p^r}$.  Analogously, consider function $f_y$ defined by $ f_y(b)=f(0,b),$ which hides $H_y$ in $\Z_{q^s}$.  The solution of the HSP in the abelian groups $\Z_{p^r}$ and $\Z_{q^s}$ with oracle functions $f_x$ and $f_y$ determines generators for the subgroups $H_x$ and $H_y$, respectively.  These subgroups have the form $H_x=\sVEV{x^{p^i}}$ and $H_y=\sVEV{y^{q^j}}$, for some $0\leq i\leq r$ and $0\leq j\leq s$.  From now on we assume that $i$ and $j$ are known.  If $j\geq t$ then $H=\sVEV{x^{p^i}y^{q^{j}}}$ (Theorem~\ref{teo-classif}), otherwise we learn that $H=\sVEV{x^{p^i}, x^ay^{q^j}}$.  In the last step we run the algorithm of Section~\ref{polynomial} to find the value of $a$ in polynomial time and with high probability.

We end this section by analyzing the time complexity of \textit{classical} algorithms for solving the HSP in $G_t$. It follows from Theorem~\ref{teo-classif} that $G_t$ has $\Omega(p^r)$ subgroups. Therefore, the HSP cannot be solved efficiently by a classical computer by performing an exhaustive search for the subgroups of $G_t$.  The methods known in the literature, such as the ones presented in Refs.~\cite{Ivanyos2,Ivanyos5} for groups with commutators of polynomial size and for nilpotent groups with constant nilpotency class, cannot be employed in this context. A remaining method is the following. The HSP in $G_t$ can be efficiently solved by finding two distinct elements $g_1$ and $g_2$ in $G_t$, such that $f(g_1) = f(g_2)$.  Let us show that such collision solves the HSP.  For each $0\leq j\leq s$, function $f$ is promised to hide subgroup $H=\sVEV{x^a y^{q^j}}$.  Then, if we know two elements $g_1$ and $g_2$ such that $f(g_1) = f(g_2)$, we will obtain
$g_2^{-1}g_1\in H$.  Using that $g_2^{-1}g_1=x^uy^v$ for some $u\in \Z_{p^r}$ and $v\in \Z_{q^s}$, we have that $g_2^{-1}g_1\in H$ if and only if
\begin{equation}\label{collision01}
\left\{
\begin{array}{l}
u\equiv a\frac{\al^{kq^j}-1}{\al^{q^j}-1} \, \up{mod}\; p^r \\
v \equiv kq^j \,\up{mod}\; q^s,
\end{array}
\right.
\end{equation}
for some $k=0,\ldots,q^{s-j}-1$.  From Eq.~(\ref{collision01}), it follows that
%\label{collision02}
$a\equiv u\frac{\al^{q^j}-1}{\al^{v}-1}\,\up{mod}\,p^r \Leftrightarrow q^t\nmid v.  $
Now the question is: What is the probability that $q^t\nmid v$ is true?  Suppose that
$v$ is an integer multiple of $q^t$, this is, $q^t \mid v$.  Since $v\in \Z_{q^s}$, there exist $q^{s-t}$ integer multiple of $v$ in $\Z_{q^s}$. The probability of $v$ being in $\Z_{q^s}$ and being an integer multiple of $q^t$ is $ \frac{q^{s-t}}{q^s}=\frac{1}{q^t}$.  Then the probability of $v$ being in $\Z_{q^s}$ and $q^t\nmid v$ is $ 1-\frac{1}{q^t}.$ With probability $1-\frac{1}{q^t}\thickapprox 1$, the HSP in $G_t$ reduces itself to the problem of finding elements $g_1\neq g_2$ such that $f(g_1)=f(g_2)$.  The problem of finding distinct elements $g_1$ and $g_2$ such
$f(g_1)=f(g_2)$ is known as the {\it collision problem}.  In that case, the function $f$ is said to be $q^{s-j}$-to-one\footnote{A function $f:X\rightarrow Y$ is said to be $m$-to-one, when there are $m$ elements in $X$ that are mapped to the same element in $Y$.}, and the time complexity of the classical algorithm for this problem is $\Theta(\sqrt{p^rq^j})$, see Ref.~\cite{samuel}.  Therefore, the lower limit of the classical algorithm for the HSP in $G_t$ is $\Omega(\sqrt {p^r})$.

\subsection{Case $H=\sVEV{x^ay^{q^j}}$ }\label{polynomial}

In this section, we present an efficient quantum algorithm that
solves the HSP in $G_t=\Z_{p^r}\rtimes\Z_{q^s}$ when $p^r/q =
\up{poly}(\log p^r)$.

The HSP in $\Z_{p^r} \rtimes\Z_{q^s}$ can be reduced to the
problem of finding cyclic subgroups generated by $x^ay^{q^j}$, 
which has order $q^{s-j}$.  We
describe a procedure that, given a function $f$ that hides the
subgroup $H=\sVEV{x^ay^{q^j}}$ in $\Z_{p^r} \rtimes\Z_{q^s}$,
efficiently determines the value of $a$ with high probability,
when $t=1$. For $t>1$, we argue that there is no efficient
solution. The procedure is the following one:

\begin{enumerate}

\item Initialize the quantum computer in the state
\begin{equation}
\sket{\Psi_1}=\frac{1}{\sqrt{q^{t-j}
p^r}}\sum_{m=0}^{p^r-1}\sum_{n=0}^{q^{t-j}-1}\sket{m}\sket{n}\sket{f(x^my^n)}.
\end{equation}
The arithmetical operations in the first ket (second ket) are
performed modulo $p^r$ $(q^{t-j})$.  Note that the left cosets of
$H$ are
\begin{equation}
x^{m_0}y^{n_0} H = \left\{ x^{{m_0} + a\al^{n_0} S(n)}
y^{n_0+nq^j}, n=0,\ldots, q^{s-j}-1 \right\},
\end{equation}
for each ${m_0}\in \Z_{p^r}$, ${n_0}\in \Z_{q^j}$  and
\begin{equation}\label{injetiva}
S(n)=\frac{\al^{nq^j}-1}{\al^{q^j} -1} \mod p^r.
\end{equation}

\item Measure the third register of state $\sket{\Psi_1}$ in the computational basis. The state after the measurement is
\begin{equation}
\sket{\Psi_2}=\frac{1}{\sqrt{q^{t-j}}}\sum_{n=0}^{q^{t-j}-1}\sket{{m_0}+a\al^{n_0}S(n)}\sket{n_0+nq^j},
\end{equation}
for some $0\leq m_0 <p^r$ and $0\leq n_0 <q^{t-j}$ unknown and
uniformly distributed. We discard the third register from now on,
because it is not relevant in what follows.

\item Apply the Fourier transform $F_{\Z_{p^r}}\otimes I$ to the state $\sket{\Psi_2}$.  The result is
\begin{equation}
\sket{\Psi_3}=\frac{1}{\sqrt{q^{t-j}p^r}
}\sum_{k=0}^{p^r-1}\sum_{n=0}^{q^{t-j}-1}\omega_{p^r}^{k({m_0}+a\al^{n_0}
S(n))}\sket{k}\sket{n_0+nq^j},
\end{equation}
where $\omega_{p^r}$ is to $p^r$-th primitive root of the unity.
\item Measure the first register in the computational basis.  Assume that the result of
the measurement is some element $k_0 \in \Z_{p^r}^{*}$.  Then, the state after the measurement is
\begin{equation}
\sket{\Psi_4}=\frac{1}{\sqrt{q^{t-j}}}\sum_{n=0}^{q^{t-j}-1}\omega_{p^r}^{k_0({m_0}+a\al^{n_0}
S(n))}\sket{k_0}\sket{n_0+nq^j}.  \end{equation} The probability
of obtaining the state $\sket{\Psi_4}$ is $1-\frac{1}{p}$.

\item Apply the operator $U$, defined by
\begin{equation}\label{U}
U\sket{m}\sket{n} = \sket{m S(n)}\sket{n-S^{-1}\left(\frac{m
S(n)}{k_0}\right)},
\end{equation}
to state $\sket{\Psi_4}$.
\end{enumerate}

Operator $U$ is not unitary in general, because $S(n)$ is not injective in 
general. However, for $n$ in $\Z_{q^{t-j}}$, Lemma~\ref{alpha^b-1} ensures that
$S(n)$ is injective, and therefore
$U$ is unitary. We must impose $t-j>t-1$, which implies $j=0$. 
It follows from Theorem~\ref{teo-classif} that $j=0$ when $t=1$. Taking
$t=1$ and applying $U$ to $\sket{\Psi_4}$ one obtains
\begin{equation}
\sket{\Psi_5}=\frac{1}{\sqrt{q}
}\sum_{n=0}^{q-1}\omega_{p^r}^{k_0({m_0}+aS(n))}\sket{k_0S(n)}\sket{0}.
\end{equation}

%\begin{equation}
%\sket{\Psi_3}=\frac{1}{\sqrt{q^{t-j}
%}}\sum_{n=0}^{q^{t-j}-1}\sket{aS(n)}\sket{nq^j}.
%\end{equation}
%%
%%
%\item  Measure the second register in the computational basis. The
%state after the measurement is
%\begin{equation}
%\sket{aS(n_0)},
%\end{equation}
%for some element $n_0\in \Z_{q^{t-j}}$. Note that $S(n_0)$ is
%known because we know $n_0$ by the measurement process. Besides,
%we obtain $S(n_0) \neq 0$ with probability $1-1/q^{t-j}$ and
%$S(n_0)\in \Z_{p^r}^{*}$ (see Lema \ref{alpha^b-1}). So,
%multiplying $aS(n_0)$ by $S(n_0)^{-1}$ we obtain the value of $a$.
%\end{enumerate}

Our goal now is to obtain parameter $a$ present in state $\sket{\Psi_5}$.
We will use the following argument discussed in Ref.~\cite{bacon1}.
Consider the state
\begin{equation} \label{perfectstate}
\sket{\tilde{a} }=\frac{1}{\sqrt{p^r} }\sum_{j=0}^{p^r-1}\omega_{p^r}^{ja}\sket{j}.
\end{equation}
Notice that an application of the inverse Fourier transform $F_{\Z_{p^r}}^{\dagger}$
to the state $\sket{\tilde{a}} $ returns the value of $a$ with high probability.
State $\sket{\tilde{a}}$ has information about parameter $a$.  What do we learn
about $a$ measuring state $\sket{\Psi_5}$?  Is there a relation between
$\sket{\tilde{a}}$ and $\sket{\Psi_5}$?  Those questions can be answered using
the notion of \textit{quantum fidelity}.

The fidelity between the quantum states $\sket{\tilde{a}}$ and $\sket{\Psi_5}$
(discarding ket $\sket{0}$ of $\sket{\Psi_5}$) is given by $\left|\langle \tilde{a}\sket{\Psi_5}\right|=\sqrt{\frac{q}{p^r}}$.
Then, applying the inverse Fourier transform $F_{\Z_{p^r}}^{\dag}$ to state $\sket{\Psi_5}$
and afterward measuring the result in the computational basis, we obtain the value of $a$
with probability $\left|\langle \tilde{a}\sket{\Psi_5 }\right|^2 =\frac{q}{p^r}$.  The total
success probability of obtaining the value of $a$ is $\left| 1-\frac{1}{p}\right|\frac{q}{p^r}=\frac{(p-1)q}{p^{r+1}}.$
Now, we run the algorithm $l$ times, where $ l =\frac{p^{r+1}}{2(p-1)q}=O(\up{poly}(\log p^r)),$
to obtain the value of $a$ with probability $1/2$.

\begin{theorem}  There is a quantum algorithm that solves, in polynomial time with success probability greater
than $1/2$, the HSP in the group $\Z_{p^r}\rtimes\Z_{q^s}$ when
$p^r/q =\up{poly}(\log p^r)$, where $p$, $q$ are distinct prime
numbers and $r, s$ are positive integers.
\end{theorem}

\section{Conclusions}
\label{conclusao}

We have presented a quantum algorithm in polynomial time for solving the HSP in a class of noncommutative groups $\Z_{p^r} \rtimes\Z_{q^s}$, where $p$, $q$ are distinct prime numbers and $r, s$ are positive integers.  Using the classification of the subgroups of $\Z_{p^r} \rtimes\Z_{q^s}$, we have showed that the HSP can be reduced to the problem of finding cyclic subgroups.  The algorithm has success probability greater than $1/2$ and requires that $p^r/q = \up{poly}(\log p^r)$ and $t=1$. For $t>1$, it seems that there is no unitary operator that reveals the parameters that describe the hidden subgroup. 

This work generalizes previous results. It generalizes the results of Ref.~\cite{MRRS} for $p$-hedral groups and Ref.~\cite{bacon1}, which are obtained from our results by setting $s=1$. It generalizes the results of Ref.~\cite{DRC1}, which are obtained from our results by setting $r=1$. In Ref.~\cite{MRRS}, the authors employed the nonabelian Fourier transform. It would be interesting to analyze the possibility of obtaining similar results for the HSP in $\Z_{p^r} \rtimes\Z_{q^s}$ using nonabelian Fourier transforms.

%\bibliographystyle{sbc}
%\bibliographystyle{hunsrt}
%\bibliography{sbc-template}
%\bibliography{weciq2008D}

\newpage

%%%%%%%%%%%%%%%%%%%%%%%%%%%%%%%%%%%%%%%%%%  AP\^{E}NDICE %%%%%%%%%%%%%%%%%%%%%%%%%%%%%%%%%%%%%%%%%%%%%%%%
\appendix
\section{Proof of Theorem~\ref{teo-classif}}
The proof of Theorem~\ref{teo-classif} uses two lemmas. The first one characterizes the cyclic subgroups of $G_t$.
\begin{lema}\label{subgroupsciclicos}
The cyclic subgroups of $G_t$ are
\begin{enumerate}
\item [i)]$\sVEV{x^ay^{q^j}}$, for $0\leq a < p^r$ and $0\leq j < t$.
\item [ii)]$\sVEV{x^{p^i}y^{q^{j}}}$, $0\leq i \leq r$ and $j \leq t\leq s.$
\end{enumerate}
\end{lema}
\begin{demo}
Let $H$ be a cyclic subgroup of $G_t$. Then $H=\sVEV{x^ay^b}$ for some $a\in \Z_{p^r}$ and $b\in \Z_{q^s}$. Let $p^i=\up{gcd}(a,p^r)$ for some $i=0,\ldots, r$ and $q^j=\up{gcd}(b,q^s)$ for some $j=0,\ldots,s$. Then, there are integers $u\in \Z_{p^r}^{*}$ and $v\in \Z_{q^s}^{*}$ such that $a=up^i$ and $b=vq^j$, respectively.  There are two cases to consider with respect to parameter $j$: $0\leq j <t$ and $ j\geq t$.  In the first case we have
$ (x^ay^b)^{v^{-1}}=(x^ay^{vq^j})^{v^{-1}}=x^{\frac{a(\al^{q^j}-1)}{\al^{b}-1} }y^{q^j} \in\sVEV{x^ay^b}.$
Notice that there is $v^{-1}$  such that $vv^{-1}=1$. Then $v\in \Z_{q^s}^{*}$.  Taking $ a' =\frac{a(\al^{q^j} -1)}{\al^{b}-1} $ we have that $x^{a' }y^{q^j}\in\sVEV{x^ay^b}$.  Since $\up{ord}({x^ay^b})=\up{ord}(x^{a' }y^{q^j})=q^{s-j}$, we have $\sVEV{x^{a' }y^{q^j} }=\sVEV{x^ay^b},$ whenever $q^j\nmid b$, for every $0\leq j <t$. Then $H$ is in class i) of Theorem~\ref{teo-classif}.  If $j\geq t$, then $ (x^ay^b)^k=(x^{up^i}y^{vq^j})^k=x^{up^ik}y^{vq^jk}=e\Leftrightarrow k = p^{r-i}q^{s-j}.  $ Besides $x^{p^i}, y^{q^j}\in\sVEV{x^{p^i}y^{q^j}} $, then $\sVEV{x^{up^i}y^{vq^j}} \subset\sVEV{x^{p^i}y^{q^j}} $.  Since $\Big|\sVEV{x^{p^i}y^{q^j} }\Big|=\Big|\sVEV{x^{up^i}y^{vq^j} }
\Big|=p^{r-i}q^{s-j}$ we have that $\sVEV{x^ay^b}=\sVEV{x^{up^i}y^{vq^j}} =\sVEV{x^{p^i}y^{q^j}} $. Then, we conclude that $H$ is in class ii).
\end{demo}

\begin{lema}\label{alpha^b-1}
Let $\alpha$ be the homomorphism that defines $G_t$. For all $b\in \Z_{q^s}$ such that $q^t\nmid b$ we have $\alpha^b-1 \in \Z_{p^r}^{*}$.
\end{lema}
\begin{demo}
Suppose by contradiction that $\alpha^b-1 \not\in  \Z_{p^r}^{*}$. Then, there is an integer number $k\in \Z_{p^r}$ such that $\alpha^b=kp+1$. Using the binomial expansion, we obtain $\al^{bp^{r-1}}=(kp+1)^{bp^{r-1}}= 1 \;\up{mod}\;p^r.$
Since ord$(\alpha)=q^t$, we have $q^t\mid bp^{r-1}$. This is a contradiction, since $q^t\nmid b$ and $p, q$ are distinct primes. Then
$\alpha^b-1 \in \Z_{p^r}^{*}$.
\end{demo}
\\
{\bf Proof of Theorem~\ref{teo-classif}} Let $H$ be a subgroup of $G_t$. If $H$ is cyclic,  Lemma~\ref{subgroupsciclicos} states that $H$ is either in class i) or in class ii) with $i=r$. Suppose that $H$ has a generating set with $n$ elements, where $n$ is a positive integer:
$H=\sVEV{x^{a_1}y^{b_1},\ldots, x^{a_n}y^{b_n}}$. If $q^t\mid b_k$ for all $k=1,\ldots,n$, let  $p^{i_k}=\up{gcd}(a_k,p^r)$ and $q^{j_k}=\up{gcd}(b_k,q^s)$, with $u_k\in \Z_{p^r}^{*}$, $v_k\in \Z_{q^s}^{*}$, for all integer numbers $0\leq i_k \leq r$ and $t\leq j_k\leq s$. Then
$
H=\sVEV{x^{p^{i_1}}y^{q^{j_1}},\ldots,x^{p^{i_n}}y^{q^{j_n}}}.
$
Define $i=\up{min}\{i_1,\ldots,i_n\}$ and $j=\up{min}\{j_1,\ldots,j_n\}$. Then, for all $k=1,\ldots, n$ we have $i_k=i+i_k'$, $j_k=j+j_k'$, for some $i_k'\in \Z_{p^r}$, $j_k'\in \Z_{q^s}$. Then
$x^{p^{i_k}}=x^{p^{i+i_k'}}=(x^{p^i})^{p^{i_k'}}\in \sVEV{x^{p^i}}$ and
$y^{q^{j_k}}=y^{q^{j+j_k'}}=(y^{q^j})^{q^{j_k'}}\in \sVEV{y^{q^j}}$.
This result implies that $x^{p^{i_k}}y^{q^{j_k}}\in \sVEV{x^{p^i},y^{q^j}}$. Since  $\sVEV{x^{p^i},y^{q^j}}=\sVEV{x^{p^i}y^{q^j}}$, we have $x^{p^{i_k}}y^{q^{j_k}}\in \sVEV{x^{p^i}y^{q^j}}$. Then $H\subset \sVEV{x^{p^i}y^{q^j}}$. Note that $x^{p^{i_k}},y^{q^{j_k}}\in H$ for all $k=1,\ldots,n$. Then $x^{p^{i}}y^{q^{j}}\in H$ and $H=\sVEV{x^{p^i}y^{q^j}}$. We conclude that $H$ is in class i). On the other hand, if $q^t\nmid b_k$ for all $k=1,\ldots,n$, then the generators of $H$ can be written as  $x^{a_k}y^{v_kq^{j_k}}$, where $v_k\in \Z_{q^s}^{*}$ with $0\leq j_k< t$. Then, for each $k,l=1\ldots,n$ with $k\neq l$, the commutator of the generators $x^{a_k}y^{b_k}$ and $x^{a_l}y^{b_l}$ is
\begin{equation}\label{maincomutator}
\Big [x^{a_k}y^{b_k}, x^{a_l}y^{b_l}\Big ]=x^{a_k+a_l\al^{b_k}-a_k\al^{bl}-a_l}=x^{\gamma_{kl}p^{i_{kl}}},
\end{equation}
for some $\gamma_{kl}\in \Z_{p^r}^{*}$ and $0\leq i_{kl}\leq r$. Let
$
i=\up{min}\{i_{kl},\;k,l=1,\ldots,n\;\up{e}\;k\neq l\}
$
and suppose with no loss of generality that $\up{ord}(x^{a_n}y^{b_n})=q^{s-j_n}\geq \up{ord}(x^{a_k}b^{b_k})=q^{s-j_k}\Rightarrow j_k \geq j_n$, for all $k=1,\ldots,n-1$. Then, we state that $H=\sVEV{x^{p^i},x^{a_n}y^{b_n}}$. In fact, it is easy to conclude that $\sVEV{x^{p^i},x^{a_n}y^{b_n}}\subset H$. We simply need to verify that $x^{a_k}y^{b_k}\in \sVEV{x^{p^i},x^{a_n}y^{b_n}}$. In fact, if $x^{a_k}y^{b_k}$ is in subgroup $\sVEV{x^{p^i},x^{a_n}y^{b_n}}$, then there are positive integers $M$ and $N$ such that
\begin{equation}
x^{a_k}y^{b_k}=x^{Mp^i+a_n\frac{\al^{Nb_n}-1}{\al^{b_n}-1}}y^{Nb_n}.
\end{equation}
This above equation implies that
\begin{eqnarray}
\left\{
\begin{array}{clc}
a_k  = &  Mp^i+a_n\frac{\alpha^{Nb_n}-1}{\alpha^{b_n}-1}&\up{mod}\; p^r; \\
b_k  = & Nb_n  & \up{mod}\; q^s. \\
\end{array}
\right.
\end{eqnarray}
This system of modular equations has solutions, since $b_k=Nb_n\;\up{mod}\;q^s\Rightarrow v_kq^{j_k}=Nv_nq^{j_n}\;\up{mod}\;q^s\Rightarrow N=v_kv_n^{-1}q^{j_k-j_n}$. To find $M$, we note that
\begin{equation}
Mp^i = a_k-a_n\frac{(\al^{b_k}-1)}{\al^{b_n}-1}.
\end{equation}
From Eq.~(\ref{maincomutator}), it follows that
$
Mp^i=\gamma_{kn}p^{i_{kn}}(1-\al^{b_n})^{-1}.
$
Using $i\leq i_{kn}$, we obtain
$
M=\gamma_{kn}p^{i_{kn}-i}(1-\al^{b_n})^{-1}.
$
Then
\begin{equation}
H=\sVEV{x^{p^i},x^{a_n}y^{b_n}}.
\end{equation}
From Lemma~\ref{alpha^b-1}, we conclude that there is the inverse $(1-\al^{b_n})^{-1}$. From Lemma~\ref{subgroupsciclicos}, the cyclic subgroup $\sVEV{x^{a_n}y^{b_n}}$ has the form $\sVEV{x^{a}y^{q^j}}$ for some $a\in \Z_{p^r}$ and some integer number $0\leq j<t$. Then
$H=\sVEV{x^{p^i},x^{a}y^{q^j}}$ is in class ii). Finally, let us suppose that subgroup $H$
can be written as $H=\sVEV{x^{a_1}y^{b_1},\ldots,
x^{a_n}y^{b_n}}$ for some $n\in \mathbb{N}$ with indices $1\leq j_1 <j_2<\ldots < j_k \leq n$ such that $q^t \nmid
b_{j_1}, \ldots, q^t\nmid b_{j_k}$. Suppose that
$j_1=1,\ldots,j_k=k$, then $H$ can be written as
\begin{eqnarray}
H & = & \sVEV{x^{a_1}y^{b_1},\ldots,
x^{a_k}y^{b_k},x^{a_{k+1} } y^{v_{k+1}q^{j_{k+1}} },\ldots,x^{a_n } y^{v_nq^{j_n}} } \\
  & = & \sVEV{x^{p^i},x^{a}y^{q^j}, x^{p^l}y^{q^m}  },
\end{eqnarray}
for some integer numbers $0\leq i,l\leq r$, $0\leq a<p^r$, $0\leq j < t $ and $t\leq m\leq s$. Define $\lambda=\up{min}\{i,l\}$, then
$
H=\sVEV{x^{p^i},x^{a}y^{q^j}, x^{p^l}y^{q^m}  }= \sVEV{x^{\lambda}y^{q^m},x^ay^{q^j}}.
$
Since $y^{q^t}=(x^ay^{q^j})^{q^{t-j}}\in \sVEV{x^ay^{q^j}}$, we have $y^{q^m}\in \sVEV{x^ay^{q^j}}$ for $m=t,\ldots,s$. This result implies that
$
H=\sVEV{x^{p^{\lambda}}, x^ay^{q^j}}.
$
Again we show that $H$ is in class ii). This ends the proof.
$\bloco$

\end{document}